\documentclass[12pt]{article}
\usepackage[cp1251]{inputenc}
\usepackage[english]{babel}
\date{}

\begin{document}

\title{Doping effect on the evolution of the pairing symmetry in n-type superconductor near antiferromagnetic phase boundary}
\author{T.\,B.\,Charikova\/\thanks{charikova@imp.uran.ru, "Low Temperature Physics", 201….}, N.\,G.\,Shelushinina, G.\,I.\,Harus, V.\,N.\,Neverov,\\ D.\,S.\,Petukhov, O.\,E.\,Petukhova,\\
{\normalsize\it Institute of Metal Physics RAS, Ekaterinburg, Russia}\\
A.\,A.\,Ivanov\\ {\normalsize\it Moscow Engineering Physics Institute, Moscow, Russia} } \maketitle

\begin{abstract}

We present the investigation results of the in-plane $\rho$(T) resistivity tensor at the temperature range 0.4-40\,K in magnetic fields up to 90kOe ($H \parallel {c}$,  $J \parallel {ab}$)  for electron-doped Nd$_{2-x}$Ce$_x$CuO$_{4+\delta}$ with different degree of disorder near antiferromagnetic - superconducting phase boundary. We have experimentally found that for optimally doped compound both the upper critical field slope and the critical temperature decrease with increasing of the disorder parameter (d-wave pairing) while in the case of the underdoped system the critical temperature remains constant and ($dH_{c2}$/$dT$)$\mid _{T_c}$ increases with increasing of the disorder (s-wave pairing). These features suggest a possible implementation of the complex mixture state as the (s+id)-pairing.


\end{abstract}
\section{Introduction}
Among the high-T$_c $ superconductors (HTSC) n-type compounds have a unique region of coexistence between the superconductivity and antiferromagnetism \cite{Motoyama2007}. The n-type compounds  Ln$_{2-x}$Ce$_x$CuO$_{4+\delta}$ (Ln=Nd, Pr, Sm) have some features compared to the p-type HTSC such as the absence of an anomalous pseudogap phase in the underdoped region of the phase diagram and the weaker electron correlations \cite{Jin2011}. Unlike the hole-doped HTSC compounds electron doping in Nd$_{2-x}$Ce$_x$CuO$_{4+\delta}$ is carried out by replacing Nd$^{3+}$ in the parent compound Nd$_2$CuO$_{4+\delta}$ with the Ce$^{4+}$ to form the non-superconducting antiferromagnet.  The annealing process removes excess oxygen and reduces the  disorder potential making the conduction electrons delocalized \cite{Ponomarev2003,Kang2007,Xu2009}.  The results of neutron scattering and scanning tunneling spectroscopy  demonstrate that antiferromagnetism (AF) and superconductivity (SC) compete locally and coexist spatially on nanometre lengh scales \cite{Zhao2011}.  The question concerning order parameter (OP) symmetry in electron-doped HTSC is under debate up to now. A number of tunneling data in optimal doped Nd$_{2-x}$Ce$_x$CuO$_{4-\delta}$ indicate weak coupling BCS-like superconductivity with s-wave OP or anisotropic s-wave symmetry in n-type superconductors \cite{Andreone1994,Kashiwaya1998,Shan2008}. The electronic Raman scattering experiments in Nd$_{1.85}$Ce$_{0.15}$CuO$_{4}$ \cite{Blumberg2002,Eremin2008} and angle-resolved photoemission (ARPES) experiments in Pr$_{0.89}$LaCe$_{0.11}$CuO$_{4}$  \cite{Matsui2005} show nonmonotonic $d_{x^{2}-y^{2}}$-type superconducting OP. There are experimental results of the magnetic penetration depth measurements in Pr$_{2-x}$Ce$_x$CuO$_{4}$ and La$_{2-x}$Ce$_x$CuO$_{4}$ \cite{Skinta2002}, point contact spectroscopy data in thin films of Pr$_{2-x}$Ce$_x$CuO$_{4}$ \cite{Biswas2002,Knobel2004}, disorder dependencies of the critical temperature T$_c$ and the upper critical magnetic field slope ($dH_{c2}$/$dT$)$\mid _{{T} \to {T_c}}$ in Nd$_{2-x}$Ce$_x$CuO$_{4+\delta}$ \cite{Charikova2009} that testify to probable transition from $d$-wave to $s$-wave or another pairing symmetry like complex mixture states ($s$+i$d$) \cite{Ruckenstein1987,Kotliar1988,Izyumov1999} with a change of the doping level. It was found (see review \cite{Izyumov1999}) that in superconductors with an anisotropic order parameter that has zeros at the Fermi surface, nonmagnetic impurities can strongly suppress $T_c$, and that the degree of suppression depends on the symmetry of the OP. So it can be used to identify this symmetry by the behavior of many properties of superconductors considered as functions of the impurity concentration. The first estimation of the behaviors of the critical temperature and the slope of the $H_{c2}$ we have made some years ago \cite{Charikova2009}, new interest have originated after fabrication of good quality underdoped single crystal films  Nd$_{1.86}$Ce$_{0.14}$CuO$_{4+\delta}$ with different degree of disorder ($\delta$) and detection of non-trivial disorder dependencies  of the  critical temperature and the slope of the $H_{c2}$ \cite{Charikova2014}. 

In this paper we present the detailed study of magnetic a.c.susceptibility and ab-magnetoresistivity of Nd$_{2-x}$Ce$_x$CuO$_{4+\delta}$ (x=0.14; 0.15) with different degree of nonstoichiometric disorder to experimentally identify the type of the pairing symmetry near the AF-SC phase transition.

\section{Experimental details}

The series of Nd$_{2-x}$Ce$_x$CuO$_{4+\delta}$/SrTiO$_3$ epitaxial films ($x$ = 0.14,  0.15) with standard (001) orientation were synthesized by pulsed laser evaporation \cite{Charikova2014}. The original target (the sintered ceramic tablet of Nd$_{2-x}$Ce$_x$CuO$_{4+\delta}$ of the given composition) was evaporated by a focused laser beam and the evaporated target material was deposited on a heated  single-crystal substrate. The substrate material was SrTiO$_3$ with (100) orientation and dimensions of 5$\times$10$\times$1.5 mm. The substrate temperature was 800$^0$\,C, the pressure during the deposition was 1.067 mbar, the residual gas was nitrous oxide (N$_2$O). Then the films were subjected to heat treatment (annealing) under various conditions to obtain samples with various oxygen content.  As a result, three types of samples with $x$=0.15 were obtained: ``as grown samples'', ``optimally reduced'' - after annealing in a vacuum at T = 780$^0$\,C for t = 60 min; p = 1.33$\cdot$10$^{-2}$\,mbar) and ``non optimally reduced'' samples - annealing in a vacuum at T = 780$^0$\,C for t = 40 min; p = 1.33$\cdot$10$^{-2}$\,mbar). Also were obtained three types of films with x=0.14: "as grown", ``optimally reduced'' - after annealing in vacuum at 60 min, 600 $^0$\,C, 1.33 10$^{-2}$\, mbar and non optimally reduced - "after annealing in vacuum" at 30,40,50 min,  600$^0$\,C, 1.33 10$^{-2}$\,mbar, films thickness was d=1600-3000\AA. Temperature dependence of the a.c. magnetic susceptibility was measured using MPMS Quantum Design SQUID magnetometer at the frequence of 81 Hz and with a.c. magnetic field amplitude of 4 Oe. Temperature dependence of resistivity were measured at the magnetic field up to 90 kOe using the Quantum Design PPMS in the temperature range $T =$ (1.8 $\div$ 300) K and the Oxford Instruments superconducting magnet up to  120 kOe in the temperature range $T =$ (0.4 $\div$ 4.2) K (Institute of Metal Physics RAS, Ekaterinburg).

\section{Experimental results and discussion}

Temperature dependencies of the resistivity $\rho$ of Nd$_{2-x}$Ce$_x$CuO$_{4+\delta}$/SrTiO$_3$ films with x = 0.14 and 0.15 and different oxygen content are presented in Fig.1(a,b). The critical temperature and the value of the resistivity depend on the annealing treatment and hence on the degree of nonstoichiometric disorder. But if in the case of optimally doped compound (x=0.15) the SC transition disappears with increase of degree of disorder (oxygen content $\delta$), in the case of underdoped system $T_c$ is practically unchanged. A similar difference is observed in temperature dependencies of the upper critical field (Fig.1(c,d)). The slope of the $H_{c2}$ decreases and tends to zero in optimally doped Nd$_{1.85}$Ce$_{0.15}$CuO$_{4+\delta}$ with increase of the degree of disorder and has no changes in the underdoped case.

The insert of the Figure 1(c) shows the a.c.susceptibility data (both $\chi'$ and $\chi''$) obtained for underdoped nonoptimally reduced single crystal film Nd$_{1.86}$Ce$_{0.14}$CuO$_{4+\delta}$/SrTiO$_3$.

The onset of diamagnetic response occurs at $T_c \simeq$ 18.8 K. At the same temperature the sufficiently sharp resistivity SC transition is observed. But there is not fairly fast drop in $\chi'$. This feature can be also seen in the $\chi''$($T$) behavior: the a.c. losses increases over the all temperature range below $T_c^{onset}$. This fact indicates the possible existence of the AF regions, where the diamagnetic response doesn't exist. However, along with this the transport percolation leads to the SC transition.

The temperature dependence of the upper critical field were determined according to the standard  resistivity method \cite{Charikova2014} in magnetic field up to $H$ = 90 kOe. 

We have measured the critical temperature ($T_c/T_{c0}$) and estimated the slope of the upper critical field $h^*$ = ($dH_{c2}/dT$)$\mid _{T_c}$/($dH_{c2}/dT$)$\mid _{T_{c0}}$. We have compared the experimental dependencies of the ($T_c/T_{c0}$) and $h^*$ on the disorder parameter $\gamma/kT_{c0}$ (Fig.2)  with theoretical calculation (Fig.2, insert) of the similar dependencies based on the microscopic derivation of Ginsburg-Landau expansion in impure system \cite{Posazhennikova1997}. In the framework of the  impurity superconductor model the disorder parameter $\gamma$ was introduced in the form:
\begin{equation}
    \gamma=\frac{h^2n_s}{4\pi m(k_F\ell)},
\end{equation}
where $n_s$ is the concentration of charge carriers in the layer,  $m$ is the electron mass and $k_F\ell$=$hc_0/\rho_{ab}e^2$ is determind experimentally due to measuring of the in-plane resistivity $\rho_{ab}$.

Figure 2 demonstrates the obtained dependencies (solid line is the theoretical dependence for the d-wave pairing, dashed curves are plotted for convenience). In insert the theoretical dependencies for $d$-wave with differnt anisotropic scattering rate and for anisotropic $s$-wave pairing \cite{Posazhennikova1997} are presented. 
In optimally doped region the critical temperature $T_c$ and the slope of $H_{c2}$ decrease with the increase of the $\gamma/kT_{c0}$ but not so abruptly as for the system with pure d-wave pairing. For the underdoped compound near the interface between AF and SC states (x= 0.14) the critical temperature remains constant with the change of the disorder parameter and the  slope of $H_{c2}$ increases with increasing of $\gamma/kT_{c0}$. Our previous investigations are shown that in overdoped regions of Nd$_{2-x}$Ce$_x$CuO$_{4+\delta}$ (x=0.18) disorder dependencies of the critical temperature and the slope of the upper critical field had the same dependencies as for anisotropic s-wave paring \cite{Charikova2012}.

Rightfulness of the study  of  $s$+i$d$ states is based on the results of antiferromagnetic model with a Van Hove singularity (AvH) \cite{Izyumov1999} at the orthorhombic distortions of a tetragonal crystal. The evidence that on the local level the symmetry of the n-doped SC is reduced with displacement of Ce was obtaned experimentally \cite{Billinge1993,Wang1990}.

\section{Conclusions}

So we have experimentally found the quite significant difference both for $T_c$($\gamma$/k$T_{c0}$)  and for the ($dH_{c2}$/$dT$)$\mid _{T_c}$($\gamma$/k$T_{c0}$) dependencies between the optimal doped  and underdoped compounds of Nd$_{2-x}$Ce$_x$CuO$_{4+\delta}$. The behavior of these values in the underdoped region  corresponds more to anisotropic $s$-wave paring and in the optimal doped - to the dirty $d$-wave type. One can to suggest that in n-type Nd$_{2-x}$Ce$_x$CuO$_{4+\delta}$ realise $s$+i$d$ pairing symmetry.

Moreover, one should take into account that the cuprates are the systems, where the antiferromagnetic state is close to the superconducting state. So the combination of the three-component magnetic OP and the two-component superconducting OP (meaning its real and imaginary parts),  as was supposed by Zhang (unified theory based on $SO$(5) symmetry) \cite{Zhang1997}, can clarify some features of the transition from the insulating to underdoped and optimally doped region.
 
This work was done within RAS Program (project N 12-P-2-1018) with partial support of RFBR (grant N 12-02-00202).

\newpage

Fig.~1. Temperature dependencies of the resistivity and the upper critical field in underdoped (a,c) and optimally doped (b,d) Nd$_{2-x}$Ce$_x$CuO$_{4+\delta}$/SrTiO$_3$ films. A.c.susceptibility at $\nu$ = 81 Hz and H = 4 Oe of underdoped nonoptimally reduced film (c, insert).

Fig.~2. The dependences of the critical temperature (a) and the slope of the upper critical field (b) on the disorder parameter for the underdoped and optimally doped Nd$_{2-x}$Ce$_x$CuO$_{4+\delta}$ compounds.

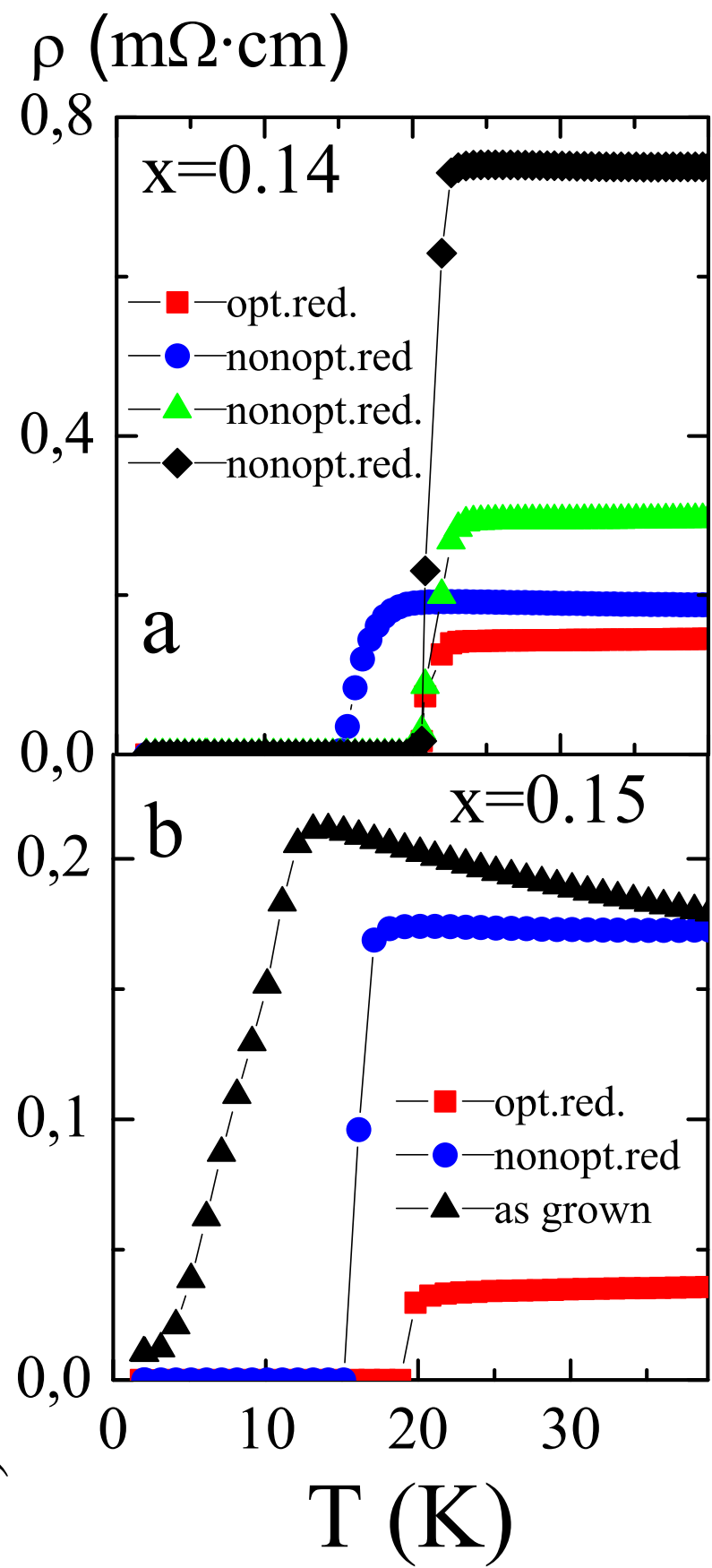
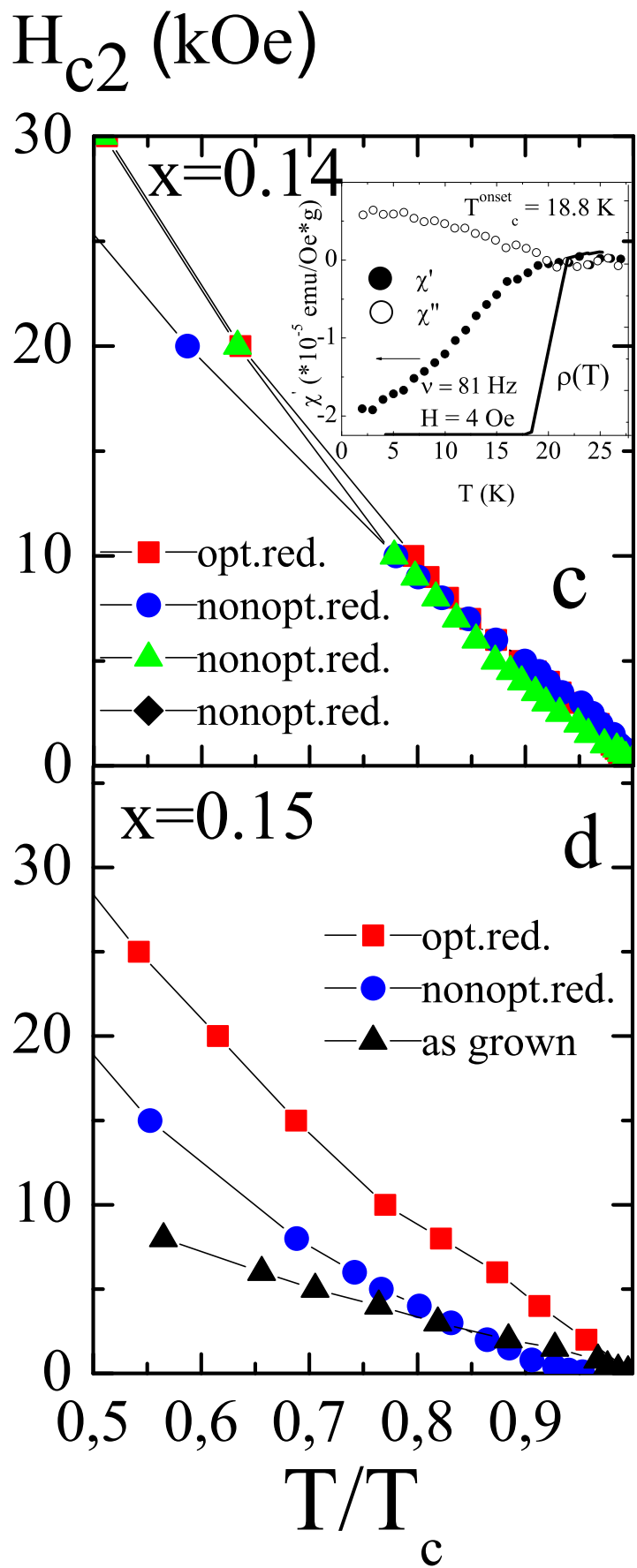
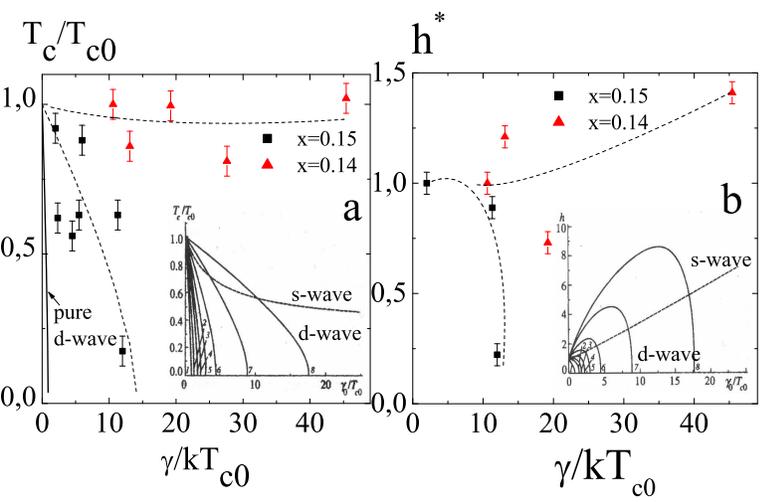
\end{document}